# Echo-processing mechanisms in bottlenose dolphins

Tengiz V. Zorikov

*Institute of Cybernetics, 5 S. Euli, 0186, Tbilisi, Georgia, zorikov@cybernet.ge*

**Abstract.** The mechanisms of echo-processing were investigated in our experiments, conducted on bottlenose dolphins. Hierarchically organized system of independent dimensions, describing echoes in animals' perception, was revealed. The rules of discrimination and recognition of echoes in dolphins were established.

## 1. Method

For exploration of the mechanisms of echo-processing in bottlenose dolphins, we developed the set of sixteen logically interconnected experiments [1-4]. Nine of them, as more informative, are presented in this paper. Other seven, mostly the control tests, were conducted to verify the validity of the gained results.

In our work, we utilized synthetic echoes produced by shock excitation of spherical piezoceramic transducer by various single and paired rectangular electrical pulses. By varying duration of electrical pulses (within the range of 6-26 μs), we were changing independently the waveforms of stimuli and consequently the coarse-scale structures of their power spectrum density (PSD). The fine-scale structures of PSD had been changing independently by varying time intervals in paired electrical pulses. As well independently, we were able to vary energy (E) and polarity of stimuli by changing amplitudes of pulses and reversing their polarity. Thus, it gave us an opportunity to make precise monitoring of dolphin's reactions to any desired physical component combined in synthetic echoes. That was reached due to the possibility of creation of any composition of components with any combinations of values of these components in synthetic echoes, – practically unattainable task for a case of real echoes.

All experiments were conducted under the same scheme. Namely, each experiment consisted of two parts. During the first one, dolphin was trained to differentiate signals of two types presented to animal alternately by series from a single transducer. Repetition rate of signals in the series was equal to 20 per second. Duration of presentation of each series was not more than 5 s. The series were presented in random order with equal probability of series type occurring. When the "positive" signals were presented, dolphin was required to move from the start position toward the transducer. The animal got a fish for the correct response. Upon presentation of the "negative" signals, the dolphin was required to remain at the start position. Once the animal had achieved stable differentiation, probe signals were presented under the same conditions, along with the "positive" and "negative" ones (the second part of experiments). Herewith, only the dolphin's swim responses to the "positive" signals were accompanied by food reward. We used other mode of reinforcement only in the last experiment. The functional role of the physical components, being analyzed in these tests, was manifested in dolphins' responses to the probe signals.

In all experiments, presented below, differences in the values of the components in the original signal pairs considerably exceeded appropriate differential thresholds in dolphins. For that reason, animals reached infallible reactions quite easily toward the end of the first parts of the tests. Therefore, we show only the second parts of these experiments on the diagrams below.

## 2. Experimental results

**Polarity**

In this test (Figure 1), we cleared up, whether or not bottlenose dolphins utilize signals' polarity in echo-discrimination tasks. The starting pair of single signals differed both in power spectrum density (PSD) and in polarity. The probe signal coincided with the "positive" one in PSD but had the opposite polarity. At the start of the second part of this test, the dolphin had equated the probe stimulus with the "positive", i.e. the animal utilized previously only PSD as the decisive factor. As it turned out, the dolphin could not distinguish between the "positive" and probe signals, identifying them throughout the test, i.e., in principle, was not able to use the differences in polarity. This result concentrated our attention on the spectral peculiarities of signals.

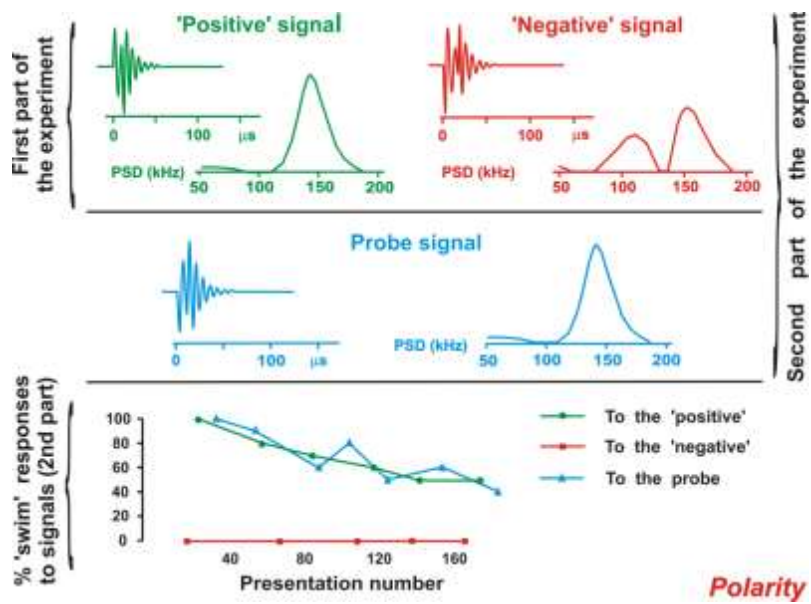

**Figure 1:** Polarity does not work as a distinctive factor. (Here and below: explanations of the experiments are contained in the text; each point of the graph represents the percentage of swim responses calculated on a set of ten subsequent presentations of the stimuli).

**Power spectrum density and energy of single pulses**

In this experiment (Figure 2), we compared PSD of a single pulses and their E. The starting pair of signals differed both in shapes of PSD and in E. Two probe signals contained crossed combinations of the values of analyzed components. The first probe signal had PSD of the "positive" stimulus and E the same as in the "negative" one. Conversely, the second probe signal coincided with the "negative" in PSD and had E the same as in the "positive". At the beginning of the second part of the test, the first probe signal had been identified by the dolphin, as the "positive" one, but the second probe signal was confidently ignored. Toward the end of the test, the unstimulated swim responses of the dolphin to the first probe signal disappeared. Thus, we can conclude that, while discriminating the initial pair of signals, the dolphin selected PSD as the decisive factor. The animal did not pay attention to difference in the energies of the "positive" and the first probe signals, having identified them at the start of the second part of the test. The dolphin had taken that component into account later, when it had become necessary for solving the task in changed conditions. This outcome demonstrated dolphin's independent perception of the analyzed components and the dominance of PSD relatively E.

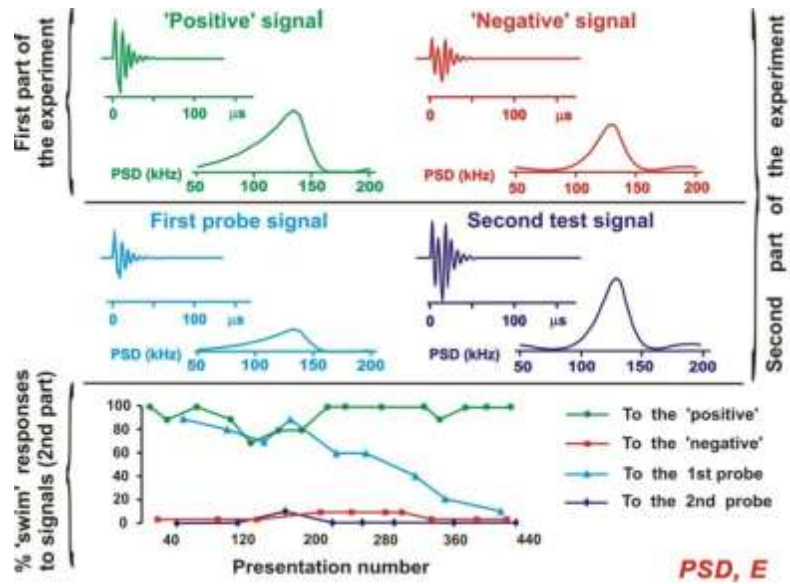

**Figure 2:** Power spectrum density (PSD) and energy (E) of single pulses are perceived independently and the first of them is dominant over the second.

**Different scale variations of PSD**

In paired stimuli, we analyzed their amplitudes and different scale oscillations of PSD. First, we compared the different scale variations of signals' PSD (Figure 3). The "positive" and the "negative" paired stimuli differed both in coarse-scale structures of PSD (envelopes of PSD) and in their fine-scale structures (periods of oscillation of PSD). Two probe signals of this test contained crossed combinations of the values of analyzed components. The first probe signal had the envelope of PSD (waveform of pulses) of the "positive" stimulus, and period of oscillation of PSD (time delay of the second highlight) the same, as in the "negative" one. Conversely, the second probe signal coincided with the "negative" in the envelope of PSD and had the period of oscillation of PSD identical to that of the "positive". At the beginning of the second part of the test, the dolphin had equated the first probe signal with the "positive" one. The second probe signal was confidently ignored. Toward the end of the test, swim responses of the dolphin to the first probe signal disappeared due to absence of rewarding. As well as above, we can conclude that, while discriminating the initial pair of signals, the dolphin preferred the envelope of PSD, as the decisive factor. The animal did not pay attention to differences in the periods of oscillation of PSD between the "positive" and the first probe signals, having equated them at the beginning of the second part of the test. The dolphin used another component later, when it had become necessary for solving the task of the second part of test. Obtained results allowed us to conclude that both components, analyzed in this experiment, are perceived by the dolphin as the independent dimensions, and the envelope or macrostructure of PSD (dimension MaPS) is dominant in animal perception over its fine-scale oscillations or microstructure of PSD (dimension MiPS).

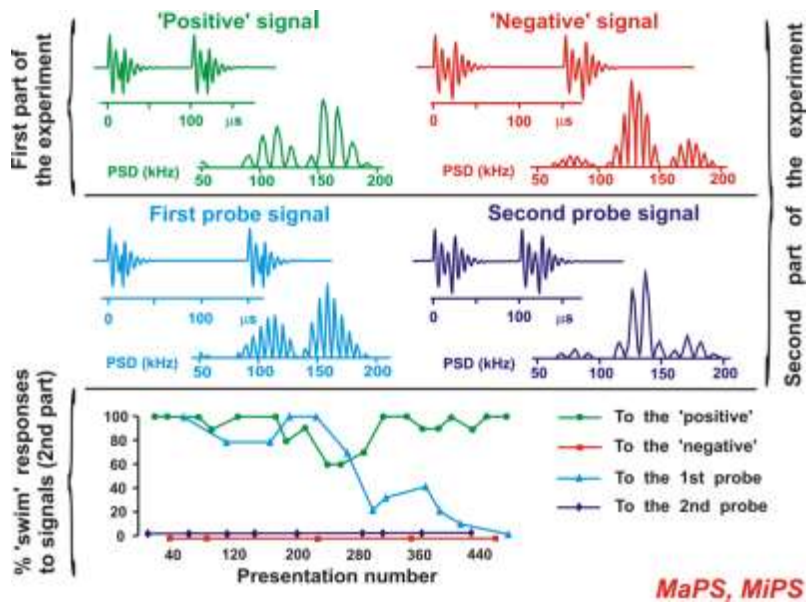

**Figure 3**: The coarse (MaPS) and fine (MiPS) spectral structures of paired pulses are perceived independently by dolphins and the first of them is dominant over the second.

**MaPS and E**

By analogy, we compared dimensions MaPS and E (Figure 4). The starting pair of signals differed in the values of both of them. Two probe signals contained their crossed combinations. At the beginning of the second part of this test, dolphin identified the first probe signal with the "positive" and ignored the second. These unstimulated swim responses disappeared toward the end of the test. The first probe signal coincided with the "positive" one in the value of dimension MaPS but differed in E. In accordance with the same logic as above, it means that both components, analyzed in this experiment, are perceived independently by the dolphin, and dimension MaPS is dominant in animal perception in relation to dimension E.

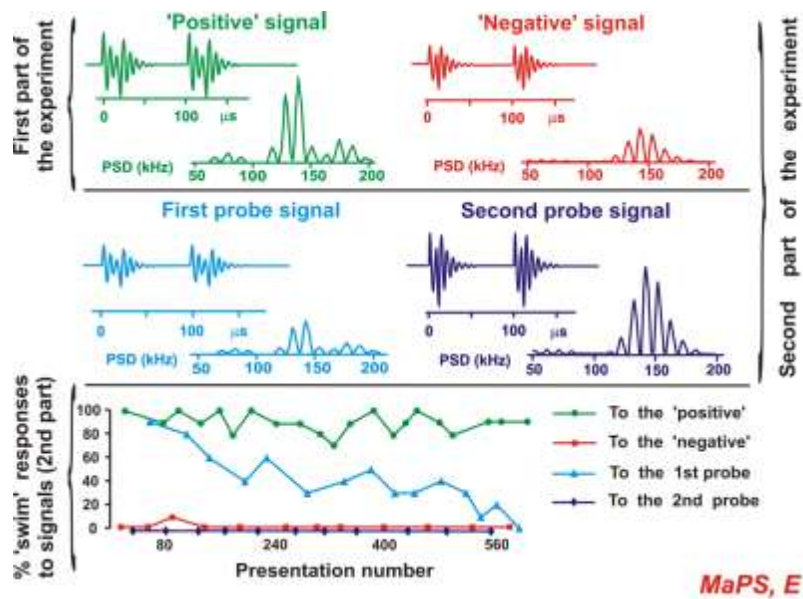

**Figure 4**: The coarse spectral structure (MaPS) and energy (E) of paired pulses are perceived independently by dolphins and the first of them is dominant over the second.

**MiPS and E**

Finally, we compared dimensions MiPS and E by the same way (Figure 4). At the beginning of the second part of this test, dolphin identified the first probe signal with the "positive", and ignored the second. As before, these swim responses disappeared toward the end of the test. The first probe stimulus coincided with the reference one in MiPS but differed in E. Accordingly, the obtained result indicates that both components, analyzed in this experiment, are perceived independently by the dolphin, and dimension MiPS is dominant in animal perception relatively dimension E.

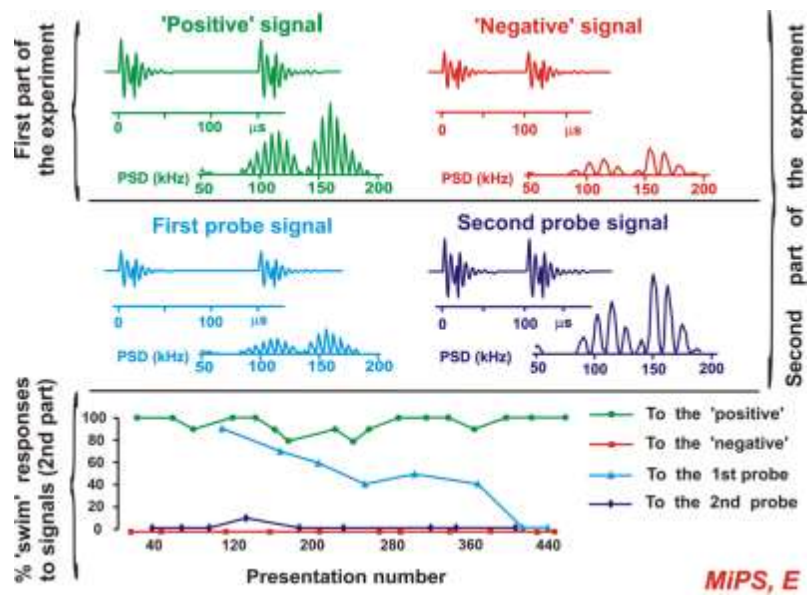

**Figure 5:** The fine spectral structure (MiPS) and energy (E) of paired pulses are perceived independently by dolphins, and the first of them is dominant over the second.

**Universal row of dimensions**

Two rows of dimensions were revealed in previous tests in single and paired stimuli: a) Signal's PSD and its E for single pulses, and b) Dimensions MaPS, MiPS, and E for the paired ones. At the same time, in accordance with the data presented above, bottlenose dolphins compare successively the values of dimensions from senior to minor, when discriminate stimuli.  Existence of one hierarchical sequence of dimensions would seem to be more adequate for such procedure.

We could eliminate this ambiguity by conducting the following experiment (Figure 6). The starting stimuli differed only by the waveforms of pulses, which constituted the pairs. The first and second probe signals of the second part of this experiment represented the single copies of the paired "positive" and "negative" stimuli, accordingly. Consequently, PSD of these probe signals coincided qualitatively with the envelopes of PSD of the "positive" and the "negative" ones. At the beginning of the second part of this experiment, the dolphin identified the first probe signal with the "positive" one. Toward the end of the test, this behavioral reaction disappeared due to lack of rewarding. The only reason for the initial identification of these signals could be the equivalence of their coarse spectral structures, or equivalence of dimensions MaPS. At the same time, the only reason for the reversing of the "positive" reaction to the first probe signal toward the end of test could be difference in their fine spectral structure, i.e. inclusion into analysis of the dimension MiPS. Thus, bottlenose dolphins utilize the same metrics to describe echoes of any structures within the CIT, i.e. one row of dimensions: MaPS, MiPS and E. This outcome eliminates the above mentioned contradiction.

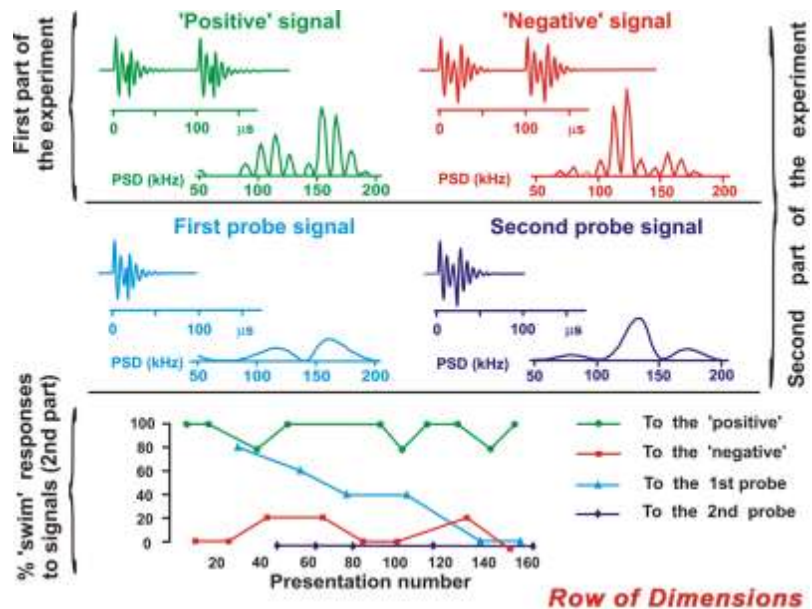

**Figure 6:** To describe echoes of any structure within the CIT, bottlenose dolphins utilize one row of dimensions: MaPS, MiPS and E (In this experiment, calculation of the percentage of swim responses fulfilled on a set of five subsequent presentations of the stimuli).

**Stability of the established hierarchical relations**

The dominance of one dimension over the other in analyzed pairs was determined by dolphins' free selection of a given dimension as the decisive factor during the first parts of the above experiments. Such hierarchical relations however could be caused by the unfortunate choice of the differences in the values of the components, being compared in the starting pairs of stimuli. It was not clear, whether the established hierarchical structure would be preserved in other relations between those differences.

Invariance of the revealed hierarchical relations was proved in the following experiments. The "positive" and "negative" signals of these experiments differed only in the values of dimensions previously established as relatively minor ones. Thereby, dolphins were intentionally compelled to accept these dimensions, as the decisive factors. In the second parts of the tests, the values of these decisive dimensions in the probe stimuli were the same, as in the "positive" signals, but the values of dimensions, earlier determined as the dominant ones in appropriate pairs, were altered. The dolphins' identification of the probe stimuli with the "positive" signals would indicate the relative nature of the established previously hierarchy. On the other hand, decision of animals to ignore these stimuli would specify that dolphins continue verifications of the values of dimensions of higher significance, despite of their evident futility in discrimination tasks of the first part of the tests.

At first, we tested stability of dominancy of dimensions MaPS and MiPS over dimension E (Figure 7, a). The starting pair of signals differed only in the values of E. First probe stimulus coincided with the "positive" in the value of dimension E, while the sole decisive factor in the first part of the test, but differed in the value of dimension MaPS. Similarly, we altered the value of dimension MiPS in the second probe signal. During the second part of this experiment, both probe signals were ignored by animal. The same result was demonstrated in the second test, in which stability of dominancy of dimension MaPS relatively dimension MiPS was tested by the same way (Figure 7, b). The starting pair

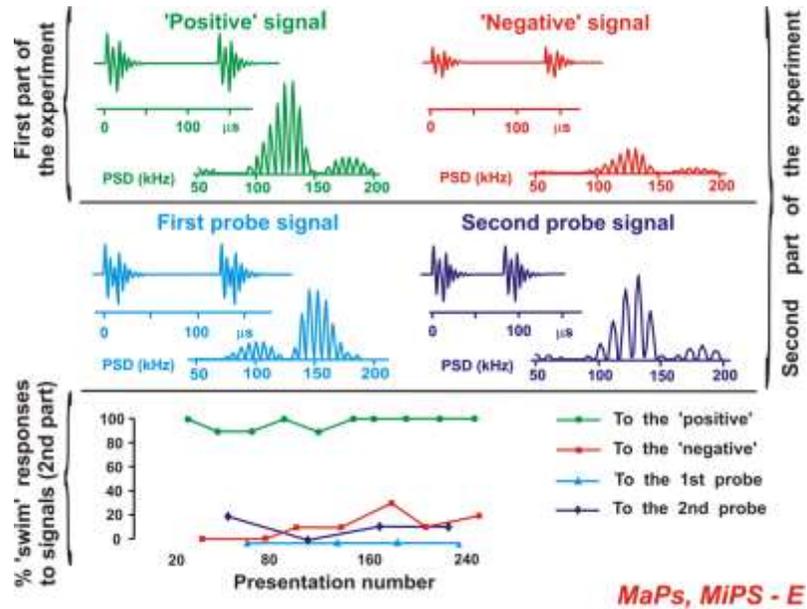

a

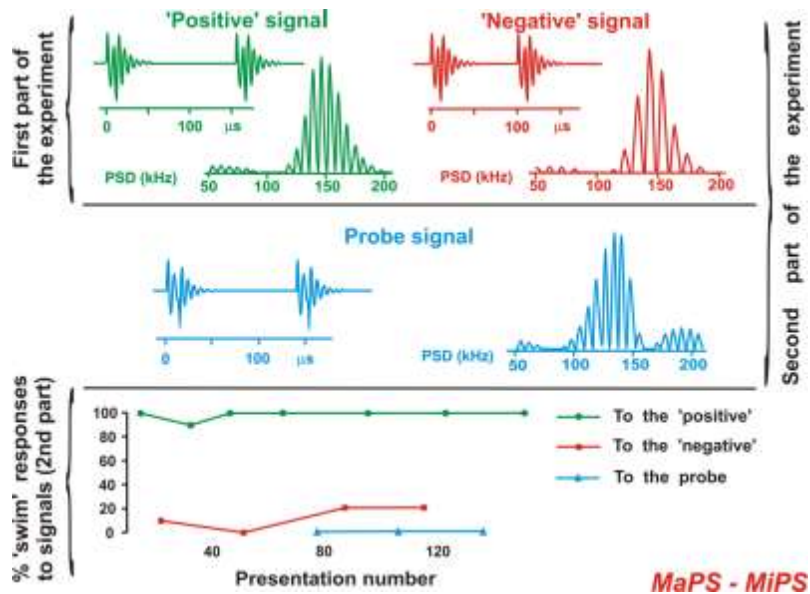

b

**Figure 7 a, b:** Hierarchical relations among the descriptive dimensions remain in force even in the limiting conditions.

of signals differed only in the value of MiPS. The probe stimulus, which coincided with the reference one in MiPS but differed in MaPS, was ignored. Thereby, the invariance of the hierarchical relations between the dimensions MaPS, MiPS and E was finally proved.

**Definitional domains of dimensions MaPS and MiPS**

It was shown at a qualitative level that different scale variations of signals' PSD govern the dimensions MaPS and MiPS. The boundaries of the definitional domains of these dimensions were assessed quantitatively in the following experiment (Figure 8). During the first part of this test, dolphin differentiated two paired signals with intervals of 120 μs between the pulses. All four pulses that constituted the pairs in the "positive" and "negative" signals had different waveforms. Paired signals, which we used as the probe stimuli in the second part of this experiment, had the same pulse structures as the "positive" one but randomly varied intervals between the pulses in the range of 50-500 μs. In this experiment, unlike the above ones, the dolphin's swim response to any of the probe stimuli was reinforced with a fish in order to prevent the animal from initiating analysis of the MiPS dimension and also to ensure equivalence of conditions for all probe stimuli. It was clear that dolphin would utilize the dominant dimension MaPS during the first part of the experiment. Thus, it was known in advance that the animal would identify, as "positive" any signal coinciding with the "positive" stimulus in the value of this dominant dimension. The probe signals were composed of pulses having the same waveforms as the pulses of the "positive" stimulus. Consequently, the value of MaPS should be the same in both cases within certain limits of variation of the interval between the pulses. This value should be lost, on the one hand, when the pulses of the probe signal would be separated by time intervals greater than the CIT. Since the merged auditory image of the pair of pulses with different time profiles within the CIT does not coincide with the separate auditory images of each of these pulses outside this interval. This interpretation should show the value of the CIT, and, at the same time, an upper boundary of the domain of definition of MiPS (in temporal expression). On the other hand, successive narrowing of intervals between the pulses (δT) in the probe signals and one-to-one related increase of the oscillation period of PSD (δF), as δF=1/δT, should inevitably distort the envelope of PSD of the pair, i.e. disrupt the value of dimension MaPS. This, in turn, may specify a definite boundary between definitional dimensions of MaPS and MiPS.

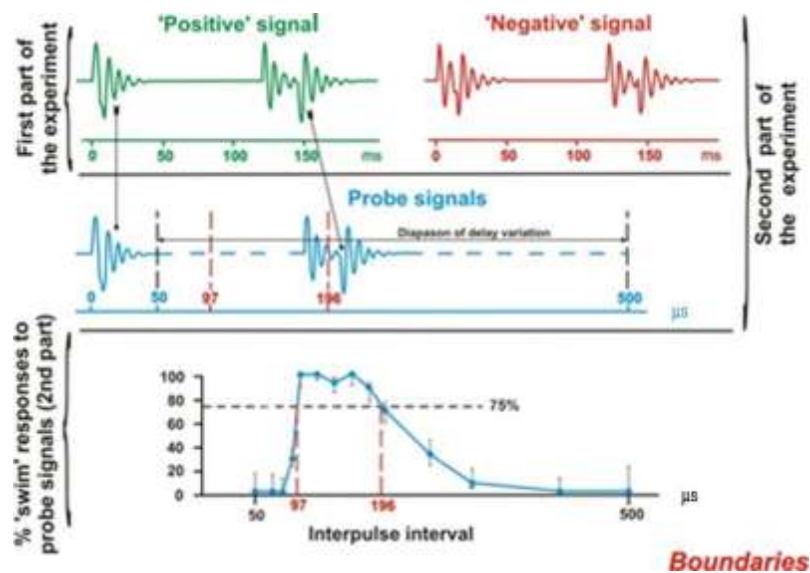

**Figure 8:** Boundaries of the definitional domains of dimensions MaPS and MiPS. Vertical line segments correspond to 95% confidence intervals.

The subsequent results confirmed the validity of the above stated hypothesis (Figure 8). The reversals of the dolphin's swim responses to the probe stimuli, caused by above-mentioned consideration, took place at the intervals of ~100 and ~200 μs (at a 75% swim responses), which, in spectral expression, correspond to the periods of oscillation of PSD ~10 and ~5 kHz accordingly. Thus, the first value characterizes the boundary between definitional domains of MaPS and MiPS. The second value represents the width of the CIT, and also the second boundary of the definitional domain of dimension MiPS.

As expected, the moment of alteration of the value of MaPS, caused by narrowing the intervals between the pulses, proves to be dependent also on the form of the envelope of PSD. In similar experiment we used shorter equiform pulses for the "positive" and probe signals, having envelopes of PSD better protected against such distortion. We obtained ~75 μs (~13 kHz, in spectral expression), instead of above stated 100 μs. This last assessment seems to be more adequate to characterize the investigated parameter.

**Conclusion**

For many years, we were exploring the sonar system of bottlenose dolphins with the goal of gathering data necessary for creation of a computer model of the echo analyzer. Altogether, six different animals were used in our tests. Our approach to the problem was rather technical, than biological, i.e. we were not interested in the constructive peculiarities of the dolphin's sonar receiver. We considered the sonar system of this animal as a "black box". Therefore, a computer model that we have created on the gained data was not an attempt of duplication of the dolphin's sonar receiver or its separate blocks but only its functional analog [3, 8 and 9].

First of all, we were interested in the process of description of echoes within the integration time window of the dolphins' sonar system, or the critical interval of time (CIT), – the time window, within which echo-highlights are transformed into a merged auditory image in animals' perception. To re-estimate this key characteristic, we used two quite different techniques (one of them is presented above). In both cases we obtained about 200 μs; considerably less value, than 265 μs that was declared in earlier papers [5, 6].

In other experiments, we have managed to show that echoes within the CIT are described with the help of three independent hierarchically interrelated dimensions, which are determined by different scales of spectral density oscillations of an echo and echo's energy. The order of domination was established, as well. Namely, the first in hierarchy is dimension MaPS, which depends on large-scale variations of echo's power spectrum that exceed ~13 kHz frequency bandwidth. The second or middle in the triad is dimension MiPS, which depends on small-scale oscillations of the echo's power spectrum with periods of ~5-13 kHz. The last in hierarchy, the minor dimension E depends on the echo's overall energy within the CIT. Apparently, we had obtained the complete system of descriptive dimensions of bottlenose dolphins' sonar, because, as it was shown in our experiment, these animals do not utilize polarity in echo-discrimination tasks.

The gained data allowed us to formulate the following rules of discrimination and recognition of signals by the bottlenose dolphins: a) Bottlenose dolphin, while distinguishing signals, compares successively the values of the descriptive dimensions from senior to minor, terminating the process at that one, which contains detectable differences in analyzed stimuli (the decisive dimension); b) If bottlenose dolphin selects some dimension as the decisive one, then in order to preserve the image of the reference signal in the animal's perception, it is necessary and sufficient to preserve the same values of the decisive dimension and all higher ones in order of hierarchy.

Apparently, such processing of echoes is realizing on the reflectory level. We obtained the similar results in experiments on humans' visual system, in which we compared perception of the shape and color [7]. We used our method of comparison, and have seen that in the initial stage of analysis dimensions color and shape are perceived independently and color is invariantly dominant in relation to shape. As well, the rules of discrimination and recognition of visual images proved to be exactly the same. However, unlike dolphins, prolongation of the first parts of those experiments up to thousands presentations had become necessary to achieve the steady reflectory level in humans' reactions.

The data, presented above, were mathematically formalized and used in our computer model. In addition, we included in the model the mechanism of averaging of dimensions' values over the series of echoes. We had obtained this result in our experiments, in which bottlenose dolphins discriminated in passive mode echoes from different actual targets, recorded beforehand separately from animals in natural conditions [1, 3]. Comparative testing of our model on synthetic and real echoes revealed critical capabilities that are not worse than those of bottlenose dolphins [3, 8, and 9].